# Numerical classical and quantum mechanical simulations of charge density wave models.

A. W. Beckwith
Department of Physics and Texas Center for Superconductivity and Advanced
Materials at the University of Houston
Houston, Texas 77204-5005, USA

**ABSTRACT**

First, using a driven harmonic oscillator model by a numerical scheme as initially formulated by Littlewood, we present a computer simulation of charge density waves (CDW); next, we use this simulation to show how the dielectric model presented via this procedure leads to a blow up at the initialization of a threshold field $E_T$. Finding this approach highly unphysical, we initiated inquiry into alternative models. We investigate how to present the transport problem of CDW quantum mechanically, through a numerical simulation of the massive Schwinger model. We find that this single-chain quantum mechanical simulation used to formulate solutions to CDW transport is insufficient for transport of soliton-antisolitons (S-S') through a pinning gap model of CDW. We show that a model Hamiltonian with Peierls condensation energy used to couple adjacent chains (or transverse wave vectors) permits formation of S-S' that can be used to transport CDW through a potential barrier. This addition of the Peierls condensation energy term is essential for any quantum model of CDW to give a numerical simulation to tunneling behavior.

**Correspondence:** A. W. Beckwith: **projectbeckwith2@yahoo.com**.

# 1. INTRODUCTION

The classical charge density wave (CDW) transport model, as presented by Gruner, answers a host of CDW questions associated with electrodynamic phenomena. However, as we show, we obtained a very non linear blow up of the calculated dielectric response of $NbSe_3$, which indicates that the Gruner model requires revision. This lead to investigations of first a single-, then a multi-chain model of CDW based upon the massive Schwinger equation model, with results we discuss herein.

Previously, we used the integral Bogomil'nyi inequality to show how a soliton-anti soliton (S-S') pair could form[1,2]. Here, we argue that this is equivalent to putting in a so called multi-chain interaction term with a constant term in it proportional to the Peierls gap times a cosine term representing interaction of different CDW chains in our massive Schwinger model,[3] which is highly unusual since at first glance adding in an additional potential energy term makes the problem look like a Josephon junction problem with no connection to the fate of the false vacuum hypothesis. We found that a single-chain simulation of the problem suffers from two defects. First, it does not answer what are the necessary and sufficient conditions for formation of a S-S'. More importantly, we also find through numerical simulations of the single-chain transport model that one needs additional physical conditions to permit barrier penetration. Our numerical simulation of the single chain problem for CDW involving S-S' gave a resonance condition in transport behavior over time, with no barrier tunneling. We argue that the false vacuum hypothesis[1,2,4] is a necessary condition for the formation of S-S' pairs — and that the multi-chain term we add to a massive Schwinger equation for CDW transport is a sufficiency condition for the explicit formation of a S-S' in our CDW transport problem. We initiate the second quantum mechanical section of this monograph by a numerical simulation of the single chain model of CDW, then show how addition of the Peierls condensensation energy permits a S-S' to form. Then, we present how to numerically simulate a multi-chain CDW simulation.

## 2. PRESENTING THE CLASSICAL WASHBOARD POTENTIAL USING LITTLEWOOD'S RANDOM PINNING MODEL

In 1986, Littlewood[5] presented an innovative scheme which incorporates a classical phase pinning model of Fukuyama, Lee, and Rice[6,7] for the interaction of impurities in a one dimensional setting. We note that, in numerical form, Littlewood's scheme bears striking semblance to the Sine-Gordon equation[8] for evolution of phase values along a one dimensional crystal. The impurity sites are randomly distributed in one dimension; we have that the phase term $\phi(x)$ represents the local 'position' of a charge density wave which interacts via an interaction potential of

$$V_j(x - R_j) \equiv V \cdot \delta(x - R_j) \tag{1}$$

which is a short range interaction between the phase $\phi(x)$ and an impurity site $R_j$. *V* turns out to be a strength of interaction term which we set equal to unity in our simulation, and the randomly chosen position of impurities, $R_j$, happens to be chosen via a random number generator in our simulation, and this selection done in such a way as to avoid bunching about certain fixed numerical quantities in a one dimensional line. This necessitated an ordering insuring $R_{j+1} > R_j$. Given this, Littlewood[5] used an overdamped equation of motion, as well as dimensionless units, in order to give an evolution equation with a first order derivative of phase with respect to time, assuming that phase $\phi(x)$ responds 'instantly' to the effects of an extremely localized interaction of phase with each impurity site given by $R_j$. This last assumption permits us to integrate between impurity sites so as to come up with a first order in time evolution equation for the phase $\phi(x_j) \equiv \phi_j$, where each $x_j \equiv c \cdot R_j$. The constant, c, is the impurity concentration and assumes that we have a correlation length L so that we observe $c \cdot L^d \gg 1$; that is, we have weak impurity pinning (here, d is the dimensionality of the spatial integration). In our model, we set d = 1.0. The remainder of this section on classical models is to look at the consequences of taking a discrete Fourier transform (DFT) of current $J \approx \langle \dot{\phi} \rangle$ to obtain computed conductivity and dielectric values due to the evolution of CDW along a one-dimensional crystal.

Several caveats are in order. First, we had to keep impurity sites from clustering too closely about the origin. When this occurs, we obtain wildly divergent computed numerical values for several computed physical quantities, especially, the derivative of phase with respect to time; this leads to spurious results for conductivity even when the applied E field is < $E_{th}$. In fact, the scheme became so unstable that, when we had numerous impurity sites near the origin, the derivative of phase with respect to time would blow up after only several dozen time steps from an initial time. In contrast with this instability, quite stable values of the derivative of phase with respect to time exist so long as the applied field to a quasi one-dimensional metal sample (e.g., NbSe$_3$) was less than a strength $V$ and applied electric field E having their dimensions *rescaled* by variable changes to non-dimensional constants. However, it is important to note that Eq. 2.1 uses a non-uniform distribution of impurity sites, when there is an interaction between phase and ions in a one-dimensional setting. However, the $\Delta\phi_i$ term in Eq. 2.1 represents the interaction between adjacent impurity sites and shows compression (or deformation) of the CDW phase, while assuming the impurity sites as given by $X_i = cR_i$ have a random distribution of $R_i$ values, while having $R_i > R_{i-1}$.

$$\dot{\phi}_i = \Delta^2 \phi_i + \frac{1}{2} E(X_{i+1} - X_i) + V \sin(\theta_i + \phi_i) \tag{2.1}$$

Eq. 2.1 is due, in part to setting the acceleration term $\ddot{\phi}$ in the *sliding condition* (uniform spacing for impurities) for CDW equal to zero (called *deep damping* due to importance of the $\frac{1}{\tau}\dot{\phi}$ term) while then, next, randomizing the position of impurity sites that is initially set equally spaced in Eq. 2.1. The $\theta_i$ expression in Eq. 2.1 is a randomized force term that varies according to a random generation of numerical values between zero and $2\pi$. Furthermore, although the sliding criteria for CDW mentioned in Eq. 2.2 assumes no spatial compression (meaning the presence of CDW only, but of no soliton), we can specifically show a distinct spatial behavior for the $\phi$ phases as generated by Eq. 2.1 above. We now refer to the uniform spacing between impurity sites equation for the evolution of phase values, by

$$\ddot{\phi} + \frac{1}{\tau}\dot{\phi} + \omega_0^2 \sin\phi = e \cdot \frac{Q}{M_F} \cdot E(t) \tag{2.2}$$

Eq. 2.1 explicitly uses $\phi_i = \phi(X_i)$ where $X_i$ = c $R_i$ and c represents impurity concentration for each impurity site on a one-dimensional line. $R_i$ represents each place on a one-dimensional line for each impurity site and is a randomly set, monotonically increasing function for each $i_{th}$ index that grows larger. We also used a discretized second derivative.[9,10,11]

$$\Delta^2 \phi_i = \frac{\phi_{i+1} - \phi_i}{X_{i+1} - X_i} - \frac{\phi_i - \phi_{i-1}}{X_i - X_{i-1}} \tag{2.3}$$

If we look at the first end point of the impurity sites, this procedure leads to a re-write of Eq. 2.3, which looks like[12]

$$\Delta^2 \phi_1 = \frac{\phi_2 - \phi_1}{X_2 - X_1} - \frac{\phi_1 - \phi_N}{X_1 - (X_N - L)} \tag{2.4}$$

where L is the grid length used in this simulation of CDW dynamics.

For the sake of including in both DC and AC contributions to an electric field, we can write

$$E = E_{dc} \tag{2.5}$$

and/or

$$\mathbf{E} = \mathbf{E}_{dc} + \mathbf{E}_{ac} \sin(\omega \tau) \tag{2.6}$$

When these electric field values are put into both Eq. 2.1, we may then examine dielectric plots which are plotted against increasing frequency according to:

$$\text{Re}\,\varepsilon(\omega) = 4\pi \left( \frac{\text{Im}\,\sigma(\omega)}{\omega} \right) \tag{2.7}$$

and

$$\text{Im}\,\varepsilon(\omega) = 4\pi \left( \frac{\text{Re}\,\sigma(\omega)}{\omega} \right) \tag{2.8}$$

$$\text{Re}\,\sigma(\omega) \propto g1 \cdot \sum_n \langle \dot{\phi} \rangle_n \cdot \cos(\omega\, t_n) \cdot \Delta t \tag{2.9}$$

as well as

$$\text{Im}\,\sigma(\omega) \propto g1 \sum_n \langle \dot{\phi} \rangle_n \sin(\omega\, t_n) \cdot \Delta t \tag{2.10}$$

As written, the derivative of phase used here is from a second-order Runge-Kutta simulation which was chosen for robustness of simulation. Having a higher-order accurate simulation for the derivative of phase, as symbolically indicated above placed in what appears to be a first-order calculation of conductivity would effectively negate the entire purpose of improved accuracy of taking the derivative of the phase calculation, as symbolically referred to in Eq. 2.1. We must perform the DFT inside the Runge-Kutta subroutine initially chosen to analyze the left side of Eq. 2.1 accurately. Otherwise, round-off error from the first-order conductivity calculation dominates, negating the second-order calculations used for the current calculation. We find that if we re-scale dielectric measurements, we re-scale dielectric measurements versus an applied electric field by resetting $\varepsilon/\varepsilon_{initial}$ in place of just $\varepsilon$ versus E field (applied to an experimental sample), and that as the frequency $\omega$ gets much smaller than $\omega_c$, we observe increasingly non-linear dielectric behavior as the E field approaches $E_{th}$. This is seen in Figures 2a, 2b,

**[Figure 2a, 2b about here]**

where we also define the critical frequency $\omega_C$ value via a convention seen in Figure 3

**[Figure 3 about here ]**

## 3. REVIEW OF THE Q.M. NUMERICAL BEHAVIOR OF A SINGLE CHAIN FOR CDW DYNAMICS.

Partly due to the failure of the classical model to avoid a blow up of the dielectric constant, addressed in the second section, we review alternate computational models that could provide some of the numerical behavior that has more overlap with known experimental features seen in previous device-development lab ( TcSAM) experiments that were performed in the 1990s up to 2000. First, we examine. a quantum mechanical

CDW model introduced by Dr. Miller, which answers certain physical issues but that we found required building in additional features.

We are modifying a one chain model of CDW transport initially pioneered by Dr. John Miller that furthered Dr. John Bardeen's work on a pinning gap presentation of CDW transport and that involves a Hamiltonian modeling how CDW would move via modeling with S-S' pairs. Qualitatively, the single-chain model is a useful way to introduce how a threshold electric field would initiate transport. We did, however, assume that the CDW would be easily modeled with a S-S' Gaussian packet, which is what we found needs further justification. With these considerations, we undertook this investigation to determine, among other things, the necessary and sufficient condition to physically justify use of a S-S' for our wave packet.

We start by using an extended Schwinger model[3,13] with the Hamiltonian set as

$$H = \int_x \left[ \frac{1}{2 \cdot D} \cdot \Pi_x^2 + \frac{1}{2} \cdot (\partial_x \phi_x)^2 + \frac{1}{2} \cdot \mu_E^2 \cdot (\phi_x - \varphi)^2 + \frac{1}{2} \cdot D \cdot \omega_P^2 \cdot (1 - \cos\phi) \right] \quad (3.1)$$

as well as working with a quantum mechanically based energy

$$E = i\hbar \frac{\partial}{\partial t} \quad (3.2a)$$

and momentum

$$\Pi = (\hbar/i) \cdot \frac{\partial}{\partial \phi}(x) \quad (3.2b).$$

The first case we are considering is a one-chain mode situation. Here, in order to introduce a time component, $\Theta \equiv \omega_D t$ was used explicitly as a driving force, while using the following difference equation due to using the Crank Nickelson[14] scheme. We should note that $\omega_D$ is a driving frequency to this physical system which we were free to experiment with in our simulations. The first index, j, is with regards to *space*, and the second, n, is with regards to *time* step. Eq. 3.3 is a numerical rendition of the massive Schwinger model plus an interaction term, where one is calling $E = i\hbar \dfrac{\partial}{\partial t}$ and one is using the following replacement

$$\phi(j,n+1) = \phi(j,n-1)$$
$$+ i \cdot \Delta t \cdot \left( \dfrac{\hbar}{D} \left[ \dfrac{\phi(j+1,n) - \phi(j-1,n) - 2 \cdot \phi(j,n) + \phi(j+1,n+1) + \phi(j-1,n+1) - 2\phi(j,n+1)}{(\Delta x)^2} \right] - \dfrac{2 \cdot V(j,n)}{\hbar} \phi(j,n) \right) \quad (3.3)$$

We use these variants of Runge-Kutta in order to obtain a sufficiently large time step interval so as to be able to finish calculations in a reasonable period of time, while avoiding an observed spectacular blow up of simulated average phase values; one so bad that one gets nearly infinite wave function values after, say 100 time steps at $\Delta t \approx 10^{-13}$. Stable Runge-Kutta simulations require $\Delta t \approx 10^{-19}$. Otherwise, one would need up to half a year on a PC in order to get the graph presented in Figure 4 below:

**[Figure 4 about here]**

A second numerical scheme, the Dunford-Frankel, which is implicit,[14] allows us to expand the time step even further. Then, the 'massive Schwinger model' equation has:

$$\phi(j,n+1) = \dfrac{2 \cdot \widetilde{R}}{1 + 2 \cdot \widetilde{R}} \cdot (\phi(j-1,n) - \phi(j+1,n)) + \dfrac{1 - 2 \cdot \widetilde{R}}{1 + 2 \cdot \widetilde{R}} \cdot \phi(j,n-1) \quad (3.4)$$
$$- i \cdot \Delta t \dfrac{V(j,n)}{\hbar} \phi(j,n)$$

where one has $\widetilde{R} = -i \cdot \Delta t \dfrac{h}{2 \cdot D \cdot (\Delta x)^2}$. The advantage of this model is that it is second-order accurate, explicit, and unconditionally stable, so as to avoid numerical blow up behavior. One then gets resonance phenomena as represented by Figure 4. This is, to put it mildly, quite unphysical and necessitates making the changes that we present in this manuscript.

## 4. ADDITION OF AN ADDITIONAL TERM IN THE MASSIVE SCHWINGER EQUATION TO PERMIT FORMATION OF A S-S' IN OUR MODEL.

Initially, we show how addition of an interaction term between adjacent CDW chains will allow a S-S' to form due to analytical considerations that we outline here. Next, we show in a numerical simulation how these terms could lead to quantum tunneling. Finally we shall endeavor to show how our argument with the interaction term ties in with the fate of the false vacuum construction of S-S' terms performed when we used the Bogomil'nyi inequality[2,15] as a necessary condition to the formation of S-S' term. Let us now first refer to how we can obtain a soliton via assuming that adjacent CDW terms can interact with each other.

There is an interesting interplay between the results of using the Bogomil'nyi inequality[2,15] to obtain a S-S' pair which we approximate via a domain thin wall approximation[2,16] and the nearest neighbor approximation of how neighboring chains interrelate with one another to obtain a representation of phase-evolution as an arctan function w.r.t. space and time variables. To whit, we can say that the Bogomil'nyi inequality provides for the necessity of a S-S' pair nucleating via a Gaussian approximation, while the interaction of neighboring chains of CDW material permits the existence of S-S' in CDW transport.

The Bogomil'nyi inequality[2,15] permits the nucleation of a S-S' pair, whereas in the argument we advance here is also pertinent whether or not we have the existence of an individual S-S'. This assumes we are using $\Delta'$ as a Peierls gap[17] energy term as an upper-bound for energy coupling between adjacent CDW chains. Note that in the argument about the formation of a S-S', we use the following equation for a multi chain simulation Hamiltonian with Peierls condensation energy[3,17] used to couple adjacent chains (or transverse wave vectors):

$$H = \sum_n \left[ \frac{\Pi_n^2}{2 \cdot D_1} + E_1[1 - \cos\phi_n] + E_2(\phi_n - \Theta)^2 + \Delta' \cdot [1 - \cos(\phi_n - \phi_{n-1})] \right] \quad (4.1a)$$

with 'momentum' we define as

$$\Pi_n = \left(\frac{\hbar}{i}\right) \cdot \frac{\partial}{\partial \phi_n} \quad (4.1b).$$

We can reverse engineer this Hamiltonian to come up with an equation of motion which leads to a soliton, via use of taking the potential in Eq. 3.1a and then use a nearest neighbor approximation to use a Lagrangian based calculation of a chain of pendulums coupled by harmonic forces to obtain a differential equation which has a soliton solution. To do this, if we say that the nearest neighbors of the adjacent chains make the primary contribution, we may write the interaction term in the potential of this problem to be[3]

$$\Delta'(1 - \cos[\phi_n - \phi_{n-1}]) \rightarrow \frac{\Delta'}{2} \cdot [\phi_n - \phi_{n-1}]^2 + \text{very small H.O.T.s} \quad (4.2)$$

and then considered a nearest neighbor interaction behavior via

$$V_{n.n.}(\phi) \approx E_1[1 - \cos\phi_n] + E_2(\phi_n - \Theta)^2 + \frac{\Delta'}{2} \cdot (\phi_n - \phi_{n-1})^2 \quad (4.3)$$

Here, we have that $\Delta' \gg E_1 \gg E_2$, so then we had a round off of

$$V_{n.n.}(\phi)\Big|_{\substack{first \\ order \\ roundoff}} \approx E_1[1 - \cos\phi_n] + \frac{\Delta'}{2} \cdot (\phi_{n+1} - \phi_n)^2 \quad (4.4)$$

which then permits us to write

$$U \approx E_1 \cdot \sum_{l=0}^{n+1} [1 - \cos\phi_l] + \frac{\Delta'}{2} \cdot \sum_{l=0}^{n} (\phi_{l+1} - \phi_l)^2 \qquad (4.5)$$

which allowed us, eventually, to obtain using $L = T - U$ a differential equation of

$$\ddot{\phi}_i - \omega_0^2 [(\phi_{i+1} - \phi_i) - (\phi_i - \phi_{i-1})] + \omega_1^2 \sin\phi_i = 0 \qquad (4.6)$$

with

$$\omega_0^2 = \frac{\Delta'}{m_{e^-} l^2} \qquad (4.7)$$

and

$$\omega_1^2 = \frac{E_1}{m_{e^-} l^2} \qquad (4.8)$$

where we assume the chain of pendulums, each of which is of length $l$ actually will lead to a kinetic energy

$$T = \frac{1}{2} \cdot m_{e^-} l^2 \cdot \sum_{j=0}^{n+1} \dot{\phi}_j^2 \qquad (4.9)$$

where we neglect the $E_2$ value. However, as we state in our derivation of the formation of a S-S' pair, having $E_2 \to \varepsilon^+ \approx 0^+$ would tend to lengthen the distance between a S-S' pair nucleating, with a tiny value of $E_2 \to \varepsilon^+ \approx 0^+$, indicating that the distance L between constituents of an S-S' pair would get very large. We did, however, find that it was necessary to have a large $\Delta'$ for helping us obtain a Sine-Gordon equation. This is so that when we set the horizontal distance of the pendulums to be $d$, the chain is of length $L' = (n+1)d$. Then, if mass density is $\rho = m_{e^-}/d$ and we model this problem as a chain of pendulums coupled by harmonic forces, we set an imaginary bar with a quantity $\eta$ as being the modulus of torsion of the imaginary bar, and $\Delta' = \eta/d$. We have an invariant quantity, which we will designate as: $\omega_0^2 d^2 = \dfrac{\eta}{\rho \cdot l^2} = v^2$, which, as n approaches infinity, allows us to write a Sine-Gordon

$$\frac{\partial^2 \phi(x,t)}{\partial t^2} - v^2 \frac{\partial \phi(x,t)}{\partial x^2} + \omega_1^2 \sin \phi(x,t) = 0 \qquad (4.10)$$

with a way to obtain soliton solutions. In order to obtain soliton solitons, we introduce dimensionless variables of the form $z = \dfrac{\omega_1}{v} \cdot x$, $\tau = \omega_1 \cdot t$, leading us to finally obtain a dimensionless Sine–Gordon equation, which we write as:

$$\frac{\partial^2 \phi(z,\tau)}{\partial \tau^2} - \frac{\partial^2 \phi(z,\tau)}{\partial z^2} + \sin \phi(z,\tau) = 0 \qquad (4.11)$$

so that

$$\phi_\pm(z,\tau) = 4 \cdot \arctan\left(\exp\left\{\pm \frac{z + \beta \cdot \tau}{\sqrt{1-\beta^2}}\right\}\right) \qquad (4.12)$$

where we can vary the value of $\phi_\pm(z,\tau)$ between 0 to $2 \cdot \pi$. Below is an example of how one can do just that: When one is looking at $\phi_+(z,\tau)$ and set $\beta = -.5$, where one has $\tau = 0$ one can have $\phi_+(z \ll 0, \tau = 0) \approx \varepsilon \approx 0$ and, also, have $\phi_+(z = 0, \tau = 0) = \pi$; whereas for sufficiently large $z$ one can have $\phi_+(z, \tau = 0) \to 2 \cdot \pi$. In a diagram with z as the abscissa and $\phi_+(z,\tau)$ as the ordinate, this soliton field from 0 to $2 \cdot \pi$ propagates with increasing time in the positive z direction and with a dimensionless velocity of $\beta$. In terms of the original variables, the soliton so modeled moves with velocity $v \cdot \beta$ in either the positive or negative *x* direction. One gets a linkage with the original pendulum model linked together by harmonic forces by allowing the pendulum chain as an infinitely long rubber belt whose width is *l* and which is suspended vertically. What we have described is a flip over of a vertical strip of the belt from $\phi = 0$ to $\phi = 2 \cdot \pi$ which moves with a constant velocity along the rubber belt. This motion is typical of the soliton we have managed to model mathematically from our potential terms above. It is very important to keep in mind the approximations used above. First, we are using the nearest neighbor approximation to simplify equation 4.4. Then, we are assuming that the contribution to the potential due to the driving force $E_2(\phi_n - \Theta)^2$ is a second order effect. All of this in its own way makes for an unusual physical picture, namely that the 'capacitance' effect given by $E_2(\phi_n - \Theta)^2$ will not be a decisive influence in deforming the solution, and is a second order effect which is enough to influence the energy band structure the soliton will be tunneling through but is not enough to break up the soliton itself.

## 5. COMPUTER SIMULATION WORK FOR MULTI CHAIN REPRESENTATIONS OF CDW TRANSPORT

Now, our Peierls gap energy[3,17] was added to the massive Schwinger equation model[13] precisely due to the prior resonance behavior with a one-chain computer simulation. We can now look at the situation with more than one chain. To do so, take a look at a Hamiltonian with Peierls condensation energy used to couple adjacent chains (or transverse wave vectors)[3,18]:

$$H = \sum_n \left[ \frac{\Pi_n^2}{2 \cdot D_1} + E_1[1 - \cos\phi_n] + E_2(\phi_n - \Theta)^2 + \Delta' \cdot [1 - \cos(\phi_n - \phi_{n-1})] \right] \quad (5.1)$$

and $\Pi_n = \left(\hbar/i\right) \cdot \partial/\partial\phi_n$ and when we will use wave functions which are

$$\Psi = N \cdot \prod_j \left(a_1 \exp\left(-\alpha \cdot \phi_j^2\right) + a_2 \exp\left(-\alpha(\phi_j - 2\cdot\pi)^2\right)\right) \qquad (5.2)$$

with a two chain analogue of

$$\Psi_{two\,chains} = N \cdot \prod_{n=1}^{2} \left(a_1 \exp\left(-\alpha \cdot \phi_j^2\right) + a_2 \exp\left(-\alpha(\phi_j - 2\cdot\pi)^2\right)\right) \qquad (5.2a)$$

If so, we put in the requirement of quantum degrees of freedom so that one has for each chain for a two dimensional case

$$|a_1|^2 + |a_2|^2 = 1 \qquad (5.3)$$

that provides coupling between nearest neighbor chains. In doing so, we are changing the background potential of this to a different situation where one has multiple soliton pairs that are due to the $\Delta'$ term in which has huge cusps given which permit the existence of tunneling due to the band structure we will present as given in Figure 5, which we will describe in the next paragraph. We will first describe a two band structure and then generalize to a five band structure we will graph in Figure 5 later on. First for a two cusp band situation with dynamical structure we have two chain interactions which we will describe here first. We should note that in tandem with NbSe$_3$ being quasi one dimensional that $\alpha \approx \dfrac{1}{\sqrt{soliton\ width}}$. For phase co-ordinate $\phi_j$, $\exp\left(-\alpha \cdot \phi_j^2\right)$ is an unrenormalized Gaussian representing a S-S' centered at $\phi_j = 0$, and a probability of being centered there given by $|a_1|^2$. Similarly, $\exp\left(-\alpha \cdot (\phi_j - 2\cdot\pi)^2\right)$. is an unrenormalized Gaussian representing a 'soliton'(anti-soliton) centered at $\phi_j = 2\cdot\pi$ with a probability of occurrence at this position given by $|a_2|^2$. We can use Eq. 5.3 to represent the total probability that one has some sort of tunneling through a potential given by Eq. 5.1 dominated by the term $\Delta'$ which dominates the dynamics we can expect due to Eq. 5.1. We then are working with

$$E(\Theta) = \langle \Psi_{two\,chains} | H_{two\,chains} | \Psi_{two\,chains} \rangle \tag{5.4}$$

We observe a band structure of sorts given by this minimum 'energy surface' given in the graph of Eq. 5.4 And we find that the term $\Delta'$ given in Eq. 4.4 is needed in order to obtain a band structure in the first place. The situation in which we have a band structure with $\Delta'(1-\cos[\phi_2 - \phi_1])$ included[3, 18] becomes complicated when we use Fortran 90, since this would ordinarily imply coupled-differential equations, which are extremely unreliable to solve numerically. For a number of reasons, one encounters horrendous round off errors with coupled-differential equations solved numerically in Fortran. Thus, when the problem was completed, instead, using Mathematica software which appears to avoid the truncation errors Fortran 90 presents us if we use a PC. with standard techniques. Here is how the problem was presented before being coded for Mathematica: where one has $E_1 = E_p =$ pinning energy, $E_2 = E_c =$ charging energy, and $\Delta' \cdot [1 - \cos(\phi_2 - \phi_1)]$ represents coupling between "degrees of freedom" of the two chains. For higher number of interacting chains, we generalize to $\Delta' \cdot [1 - \cos(\phi_n - \phi_{n-1})]$ When we had five interacting chains, the wave function was set to a different value than given in either Eq. 5.2 or Eq. 5.2a

$$\Psi_m(\phi_i) = \sum_{m=-2}^{2} b_m \exp(-\alpha(\phi_i - 2 \cdot \pi \cdot m)) \tag{5.5}$$

with:

$$\sum_{m=-2}^{2} b_m^2 = 1 \tag{5.6}$$

we obtained a minimum energy 'band structure' with five adjacent parabolic arcs[3,18]. We obtain a minimum energy out of this we can write as:

$$E = E_{min} = \langle \Psi | \hat{H} | \Psi \rangle \tag{5.7}$$

where $D_1 = 174.091, E_p = .00001, E_c = .000001$ and $\Delta' = .005$ for Hamiltonian

$$\hat{H}_{two\,chains} = \sum_{n=1}^{2} \left[ \frac{\Pi_n^2}{2 \cdot D_1} + E_1[1 - \cos\phi_n] + E_2(\phi_n - \Theta)^2 + \Delta' \cdot [1 - \cos(\phi_n - \phi_{n-1})] \right] \tag{5.8}$$

where minimum energy curves are set by the coefficients of the two wave functions, which are set as $b_{-2}, b_{-1}, b_0, b_1, b_2; c_{-2}, c_{-1}, c_0, c_1, c_2; \alpha$ (which happens to be the wave parameter for Eq. 5.6.

**[Put Figure 5 about here]**

This leads to an energy curve given in Figure 5 where there are five local minimum potential energy values. It is a reasonable guess that for additional chains (i.e. if m bracketed by numbers > 2) that the number of local minimum values will go up, provided that one uses a modified version of numerical simulation wave function as given in Eq. 5.5. We performed the following to plot an average <phi> value, which we will represent in equation 5.10. The easiest way to put in a time dependence in the Hamiltonian (Eq. 5.8) is to provisionally set $\Theta = \omega_D t$ for the graphics presented, $\omega_D = 0.67 \, \text{M Hz}$.

If we set $\Psi \equiv \Psi(\phi_1, \phi_2, \Theta)$ which has an input from the Hamiltonian $\hat{H}_{two\,chains}$

then we can set up an average phase, which we will call:

$$\Phi = \frac{1}{2}(\phi_1 + \phi_2) \tag{5.9}$$

where we calculate a mean value of phase given by[3, 18]

$$\langle \Phi(\Theta) \rangle = \int_{-\eta\pi}^{\eta\pi} \int_{-\eta\pi}^{\eta\pi} d\phi_1 d\phi_2 \frac{1}{2} \cdot (\phi_1 + \phi_2) |\Psi(\phi_1, \phi_2, \Theta)|^2 \tag{5.10}$$

The integral $\langle \Phi(\Theta) \rangle$ was evaluated by 'Nintegrate' of Mathematica, and was graphed against $\Theta$ in Figure 6, with $\eta = 20$

**[Figure 6 about here]**

These total sets of graphs put together are strongly suggestive of tunneling when one has $\Delta \neq 0$ in $\hat{H}_{two\,chains}$.

The simulation results of Figure 6 are akin to a thin wall approximation leading to a specific shape of the S-S'pair in *phase*-space, which is also akin to when we have abrupt but finite transitions after long periods of stability[1,2]. We can link this sort of abrupt transitions to what happens when we have a 'thin wall approximation ' as spoken of by Sidney Coleman in his "fate of the false vacuum" hypothesis[4] for *instanton* transitions. We do, however, need to verify if or not that the soliton solution to this problem is optimal for tunneling. Trying to show this will be the main reason for the next section treatment of how a multi-chain interaction will be a necessary condition for formation of S-S' in CDW transport problems. This is when we will be working with wave functionals of the form given by initial and final wave functionals looking like

$$\Psi_i[\phi(x)]_{\phi \equiv \phi_{ci}} = c_i \cdot \exp\left\{-\alpha \cdot \int dx [\phi_{ci}(x) - \phi_0(x)]^2\right\},$$
$$\Psi_f[\phi(x)]_{\phi \equiv \phi_{cf}} = c_f \cdot \exp\left\{-\alpha \int dx [\phi_{cf}(x) - \phi_0(x)]^2\right\}$$
(5.11a,b)

## 6: CONCLUSION: SETTING UP THE FRAMEWORK FOR A FIELD THEORETICAL TREATMENT OF TUNNELING.

We have, in the above identified pertinent issues needed to be addressed in an analytical treatment of CDW transport. First, we should try to have a formulation of the problem of tunneling that has congruence with respect to the Sidney Coleman false-vacuum hypothesis. We make this statement based upon the abrupt transitions made in a multi-chain model of CDW tunneling that are identical in form to what we would expect in a thin-wall approximation of a boundary between true and false vacuums. Secondly, we can say that it is useful to keep a S-S' representation of solutions for CDW transport. Figure 5 and Figure 6 address minimum conditions for the formation of a S-S'; what we have outlined here, however, concerns when we want to have a band structure pertinent to tunneling analysis; then we should keep the $\Delta'$ term necessitated in coupling chains together in CDW transport analysis.

We explicitly argue that a tunneling Hamiltonian based upon functional integral methods is essential for satisfying the necessary and sufficient conditions for the formation of a S-S' pair. The Bogomil'nyi inequality stresses the importance of the relative unimportance of the driving force $E_2 \cdot (\phi_n - \Theta)^2$, which we drop out in our formation of a S-S' in our multi chain calculation. In addition, we argue those normalization procedures, plus assuming a net average value of the $\Delta'(1 - \cos[\phi_n - \phi_{n-1}]) \to \frac{\Delta'}{2} \cdot [\phi_n - \phi_{n-1}]^2$ + small terms as seen in our analysis of the contribution to the Peierls gap contribution to S-S' pair formation in our Gaussian[1,2] wave functional Eqs. 5.11a,b, where the normalizing term $c_{i,f}$ would allow us to scale out an averaged out value of the $\Delta'(1-\cos[\phi_n - \phi_{n-1}]) \to \frac{\Delta'}{2}\cdot[\phi_n - \phi_{n-1}]^2$, and representation of how S-S' pairs interact in a multi-chain model evolve in a pinning gap transport problem for CDW dynamics. This would allow us, if done, to have S-S' pairs being used in equation 5.13 due to their formation in our problem and due the Peierls gap term in the Hamiltonian [3,18].

Then the fate of the false vacuum hypothesis used [1,2,3] so that the S-S' pairs can have nucleation behavior as seen in Figure 7

**[Figure 7 about here ]**

is consistent with a Gaussian wave functional representation of transport behavior [1,2] leading to matching with experimentally observed current behavior [2,3] as seen in Figure 8. This would permit us to form necessary and sufficient conditions for permitting a Gaussian wave functional to use S-S' pairs to form the current experimentally observed in Figure 8, where after a long derivation[1,2,3] we have

**[Figure 8 about here ]**

where we write

$$I \propto \tilde{C}_1 \cdot \left[\cosh\left[\sqrt{\frac{2\cdot E}{E_T \cdot c_V}} - \sqrt{\frac{E_T \cdot c_V}{E}}\right]\right] \cdot \exp\left(\frac{-E_T \cdot c_V}{E}\right) \qquad (6.1)$$

with $\tilde{C}_1 \equiv \frac{C_1 \cdot C_2}{2\cdot m^*}$

which is a significant improvement over a prior expression derived to qualitatively fit experimental data[19]

$$I \propto G_P \cdot (E - E_T) \cdot \exp\left(\frac{-E_T}{E}\right) \text{ if } E > E_T \tag{6.2}$$

otherwise

## FIGURE CAPTIONS

**Figure 1:** Average phase $\langle \phi \rangle$ plotted against time (for $E_{dc}$) with $\langle \phi \rangle$ stabilizing if $E_{dc} < E_{th}$ and $\langle \phi \rangle$ monotonically increasing if $E_{dc} > E_{th}$.

**Figure 2:** Comparison of scaled dielectric values when one has signal frequency $\omega \leq \omega_c$ i.e. near a critical value $\omega_c$. **Figure 2a** applies to low frequency plots, and **Figure 2b** to high frequency plots. One obtains the situation that there is a blow up of the dielectric response if one has the electric field exceeding a threshold value, which could not be duplicated numerically. The dielectric is infinite valued when $E=E_{th}$

**Figure 3:** This conductivity plot shows the origins of how we pick critical value $\omega_c$

**Figure 4:** Beginning of resonance phenomena in single chain quantum dynamics due to using the traditional Crank–Nickelson numerical iteration scheme of the one-chain model.

**Figure 5:** Determining band structure via a *Mathematica 8* program, with wave functions given by Eq. 4.6.

**Figure 6:** Phase vs. $\Theta$, according to the predictions of the 'multi-chain'-tunneling model.

**Figure 7:** Evolution from an initial state $\phi_i$ to a final state $\phi_f$ for a double-well potential (inset) in a 1-D model, showing a kink-antikink pair bounding the nucleated bubble of true vacuum. The shading illustrates quantum fluctuations about the classically optimum

configurations of the field $\phi_i(x) = 0$ and $\phi_f(x)$, while $\phi_0(x)$ represents an intermediate field configuration inside the tunnel barrier

**Figure 8** :Experimental and theoretical predictions of current values. The dots represent a Zenier curve fitting polynomial, whereas the blue circles are for the S-S$'$ transport expression derived with a field theoretic version of a tunneling Hamiltonian.

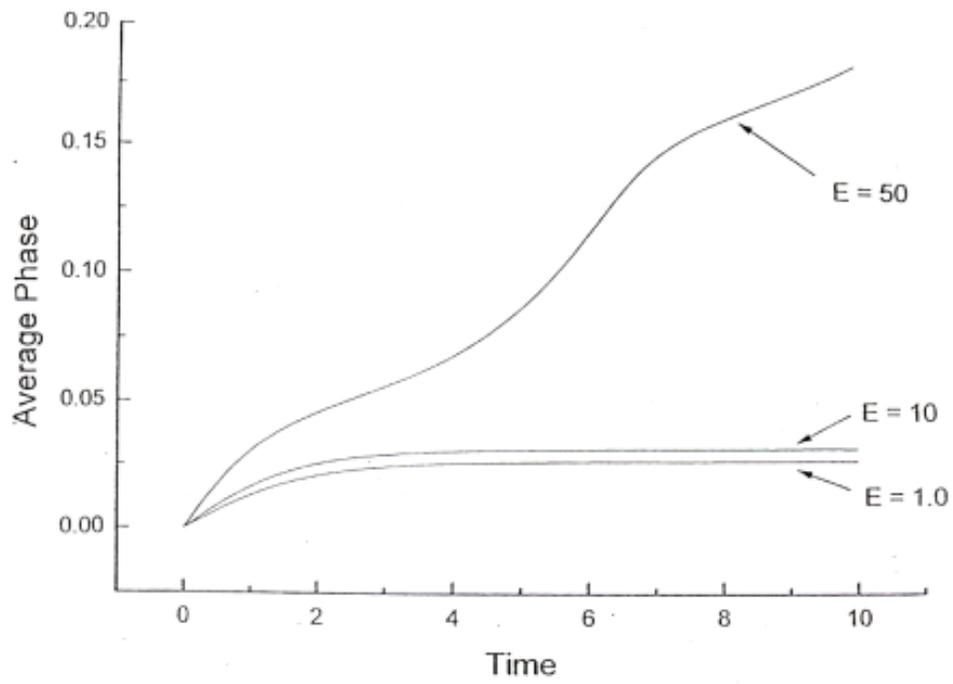

**FIG 1**
**Beckwith**

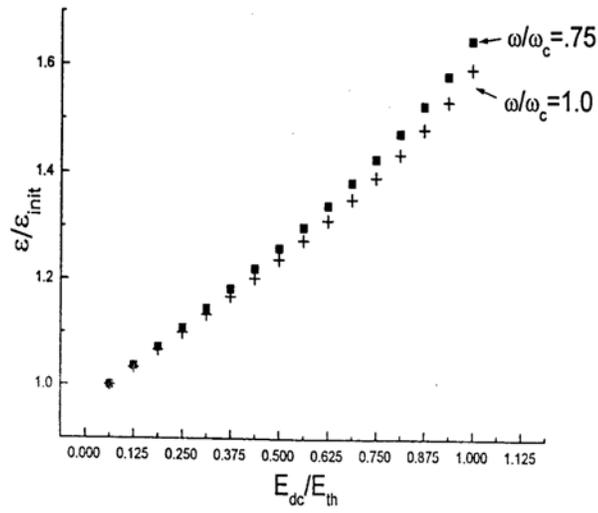

**Figure 2a**

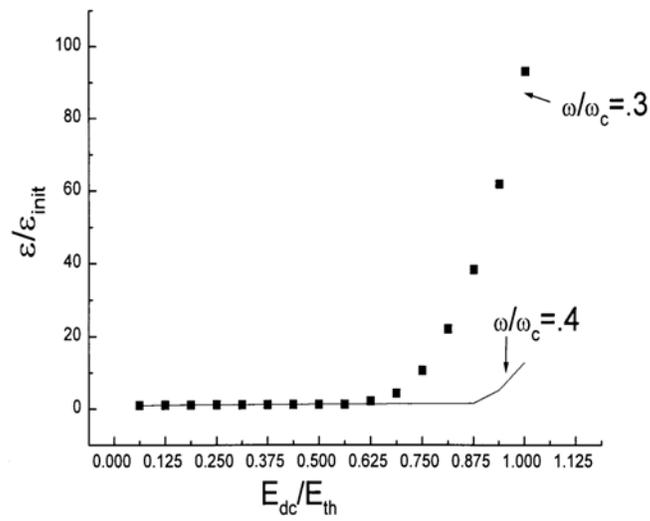

**Figure 2b**

**Beckwith**

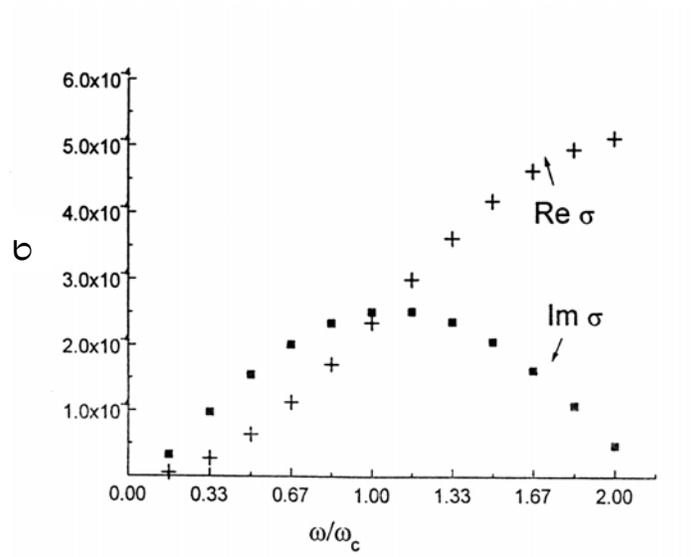

**Figure 3**

**Beckwith**

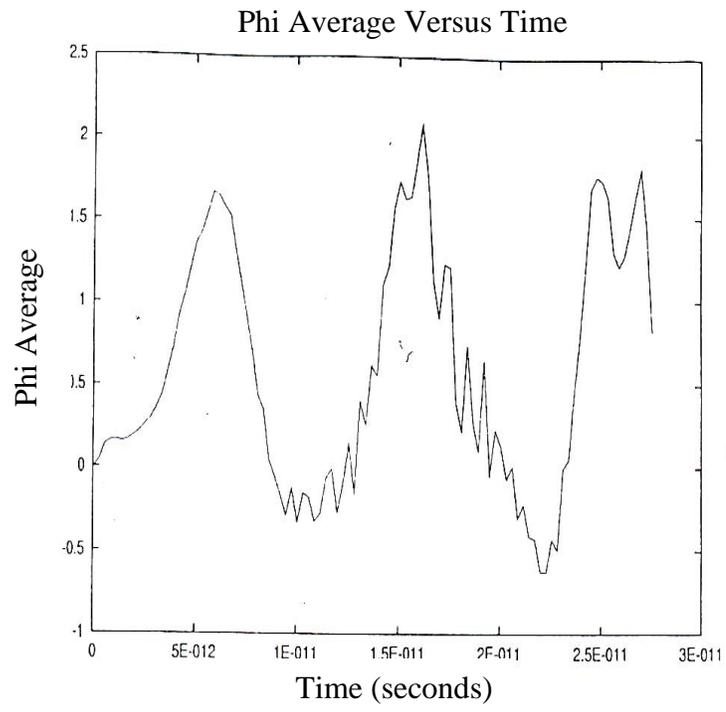

**Figure 4**

**Beckwith**

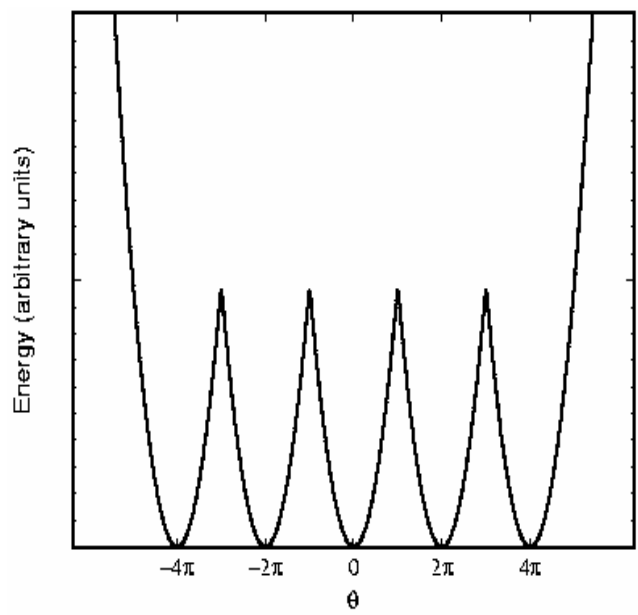

**Figure 5**

**Beckwith**

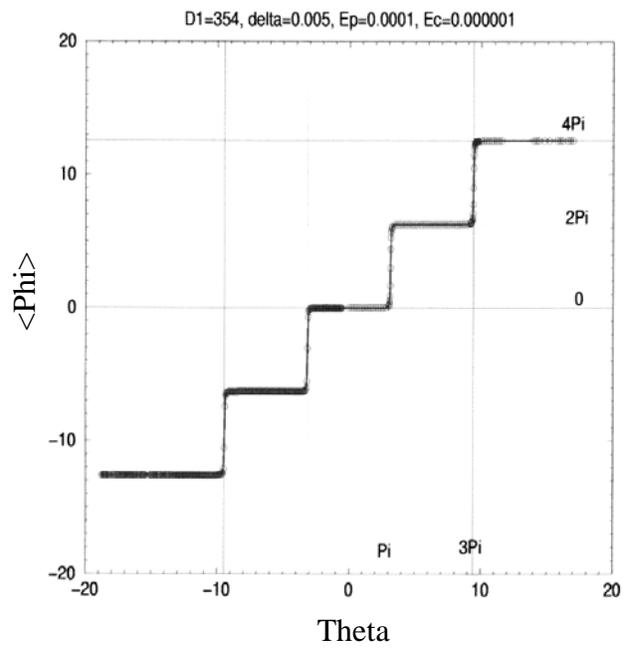

**FIGURE 6**
**Beckwith**

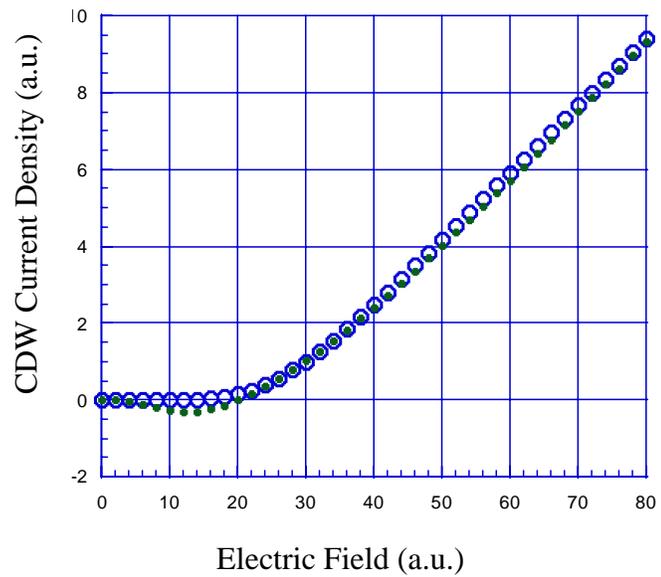

**FIGURE 7**
**Beckwith**

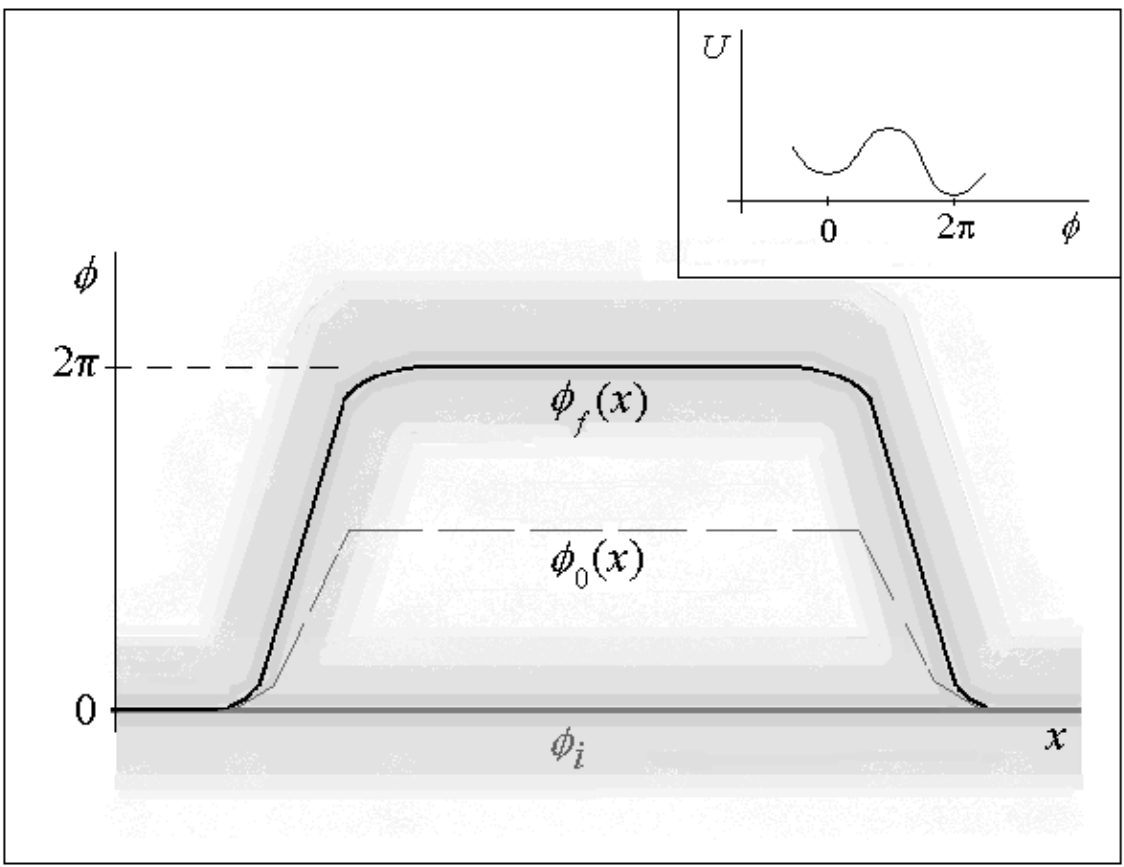

**Figure 8**

**Beckwith**